# Longitudinal Momentum Mining of Beam Particles in a Storage Ring

C. M. Bhat

Fermi National Accelerator Laboratory, P.O.Box 500, Batavia, IL 60510, USA

I describe a new scheme for selectively isolating high density low longitudinal emittance beam particles in a storage ring from the rest of the beam without emittance dilution. I discuss the general principle of the method, called longitudinal momentum mining, beam dynamics simulations and results of beam experiments. Multi-particle beam dynamics simulations applied to the Fermilab 8 GeV Recycler (a storage ring) convincingly validate the concepts and feasibility of the method, which I have demonstrated with beam experiments in the Recycler. The method presented here is the first of its kind.

One of the most important problems encountered in high-energy hadron beam storage rings is to select only the high intensity, low emittance, region of the *phase space* of beam particles with minimal emittance dilution. Considerable progress has been made over the last fifty years[1][2] on techniques broadly referred to as momentum mining, namely, the selective isolation of the beam particles in the high density, low longitudinal emittance region from the rest of the phase space. Typically, the region of interest in the beam particle distribution lies in the vicinity of the *synchronous* particle[3]. Antiproton storage rings both at CERN[4] and at Fermilab[5] have adopted what I call transverse momentum mining for extracting the dense beam from a stack of cooled beam particles. For example, in the antiproton Accumulator Ring at Fermilab, a part of the stored beam is captured adiabatically in *buckets* of sinusoidal radio-frequency (rf) waves with *h (harmonic number)*=4 and *bucket area* smaller than the total beam phase space area. Subsequently, the beam particles in *h*=4 buckets are pulled out from the main stack transversely through acceleration. Once they are completely outside of the main stack, they are extracted. The shortcoming of this method is the inevitable beam disruption caused by the separation of the low emittance particles from the beam. Consequently, this leads to longitudinal emittance growth and, if the beam extraction is carried out multiple times, the later extractions suffer from lower particle density.

Previously, a method for proton mining using dual frequency amplitude modulation has been proposed,[6,7] which is very similar to the transverse mining method.

In this paper, I propose a new technique for mining[8] beam particles from the high density region of the *longitudinal* phase space with minimal emittance dilution. The mining is done using a rectangular *barrier*[9] rf system.

A particle beam in a storage ring is characterized by its energy spread $\Delta \hat{E}$ about its synchronous energy $E_0$ and a characteristic transverse emittance. In the absence of *synchro-betatron coupling* these two quantities can be varied independently of each other. Generally, the energy distribution of the particles in the storage ring is approximately parabolic or Gaussian in shape with the synchronous particles at the peak and the particles of lower energy spread closer to the peak than those with a greater spread. The Hamiltonian of any particle with energy $\Delta E$ relative to the synchronous particle in a *synchrotron* is given by,[10,11]

$$H(\tau, \Delta E) = -\frac{\eta}{2\beta^2 E_0}\Delta E^2 - \frac{e}{T_0}\int_0^\tau V(t)dt, \qquad (1)$$

where $\eta$, $T_0$ and $\beta$ are the phase slip factor, the revolution period and the ratio of the particle velocity to that of light, respectively, and $-\tau$ is the time difference between the arrival of this particle and that of a synchronous particle at the center of the rf bucket. $V(t)$ is the amplitude of the rf voltage wave-form and $e$ is electronic charge. We identify the second term of the above equation as the potential energy $U(\tau)$ of the particles, given by,

$$U(\tau) = -\frac{e}{T_0}\int_0^\tau V(t)dt. \qquad (2)$$

For a rectangular barrier bucket, $V(t)$ is given by,

$$V(t) = \begin{vmatrix} -V_0 & \text{for } -T_1 - T_2/2 \leq t < -T_2/2, \\ 0 & \text{for } -T_2/2 \leq t < -T_2/2, \\ V_0 & \text{for } T_2/2 \leq t < T_1 + T_2/2, \end{vmatrix} \qquad (3)$$

where $T_1$ and $T_2$ denote barrier pulse width and gap between rf pulses as shown in Fig. 1. A schematic view of the rf wave form with the beam phase space boundary (dashed line in left figure) and the corresponding potential well, containing beam particles for a storage ring operating below *transition energy* is shown in Fig. 1(a).

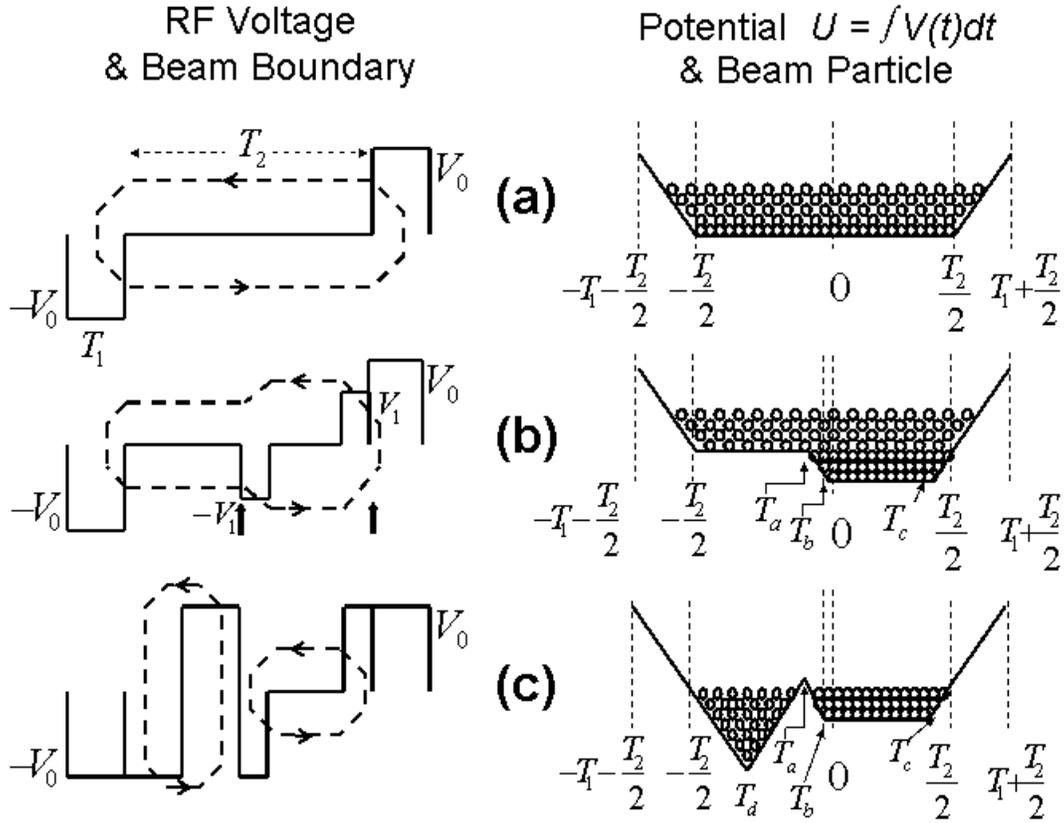

FIG. 1. Schematic view of longitudinal momentum mining using barrier buckets. Barrier rf voltage (solid-lines) and beam particle boundary in ($\Delta E$, $\tau$)-phase space (dashed line) are shown on the left. The cartoons on the right show potential well and the beam particles in it. (a) The initial distribution, (b) after confining particles with low energy spread in a deeper well and (c) after isolating particles with high and low energy spreads.

The objective of longitudinal momentum mining is to isolate particles closer to $E_0$ from the rest. This is accomplished by adiabatically inserting a mining bucket inside the existing well (between $-T_2/2$ and $T_2/2$), as indicated by arrows in Fig. 1(b), so that all particles with energy near the synchronous energy, including synchronous particles, drift to the lowest potential. It is important to note that the trapping of particles take place during the adiabatic opening of the mining bucket. Since the synchrotron oscillation periods of the particles with energy closest to $E_0$ are very large, the drift times are very long. To expedite the mining process and to ensure the trapping of particles with energy $E_0$, one can grow a negative pulse immediately to the right of the left-most rf pulse (at $-T_2/2$) and shovel adiabatically to a location to the right indicated by $T_a$. The final voltage wave form for this configuration is given by,

$$V(t) = \begin{vmatrix} -V_0 & \text{for } -T_1 - T_2/2 \leq t < -T_2/2 \\ 0 & \text{for } -T_2/2 \leq t < T_a \\ -V_1 & \text{for } T_a \leq t < T_b \\ 0 & \text{for } T_b \leq t < T_c \\ V_1 & \text{for } T_c \leq t < T_2/2 \\ V_0 & \text{for } T_2/2 \leq t < T_1 + T_2/2 \, . \end{vmatrix} \quad (4)$$

The longitudinal emittance of the trapped particles in the mining bucket between $T_a \leq t < T_2/2$ is given by $\varepsilon_m = 2(T_c - T_b)\Delta E_m + 4T_0|\eta|\Delta E_m^3/(3\beta^2 E_0 eV_1)$ where $\Delta E_m = \sqrt{2\beta^2 eV_1(T_b - T_a)E_0/(T_0|\eta|)}$ and $e$ is the electron charge. Finally, to isolate the remaining particles from those that are trapped, another rf bucket is opened in the region $-T_2/2 \leq t < T_a$ as shown in Fig. 1(c). Thus are the particles with low longitudinal emittance mined while leaving the rest in the region $-T_2/2 \leq t < T_a$.

We have applied the scheme described above to the beam in the Fermilab Recycler[12]. The Recycler is an 8 GeV synchrotron storage ring that operates below the transition energy ($\gamma_T = 21.6$) and has $T_0 = 11.12$ μsec. This will be the main antiproton source for the proton-antiproton collider programs at Fermilab. The antiproton beam is stacked and stored azimuthally in the Recycler using barrier buckets generated by a broad-band rf system[13] capable of providing rf pulses of 2 kV. The beam is cooled initially using stochastic cooling[14] and is expected to be cooled further with electron cooling[15] to ≤ 54 eVs longitudinally and ≤ 7 π-mm-mr transversely. We plan to accumulate > 6×10$^{12}$ antiprotons in the Recycler before they are transferred to the Tevatron[5]. Either a part of the cooled beam or the entire stack will be extracted in nine transfers of equal emittance and equal intensities, each with four 2.5 MHz bunches. Thus, there will be thirty six antiproton bunches in the Tevatron from the Recycler. The longitudinal emittance of each bunch at the time of extraction is expected ≤1.5 eVs. Stacking and un-stacking of antiprotons from the Recycler entails complicated sets of rf manipulations[16]. In order to achieve high proton-antiproton *luminosity* in the Tevatron, it is essential to maintain the emittance of the beam throughout the chain of rf manipulations in the Recycler and to send only a high density low longitudinal emittance beam to the Tevatron.

Our testing of the method of longitudinal momentum mining in the Recycler was carried out in two steps. First, computer simulations using a multi-particle beam dynamics code, ESME[17], were carried out to establish the mining steps. Then, experiments were done with beam in the Recycler to demonstrate the technique.

The simulation results presented here assume about 100 eVs of beam in the Recycler captured in a rectangular barrier bucket. The primary goal was to mine 54 eVs of the low longitudinal emittance high density part of the phase space distribution of the beam and capture the rest of the beam in a separate bucket.

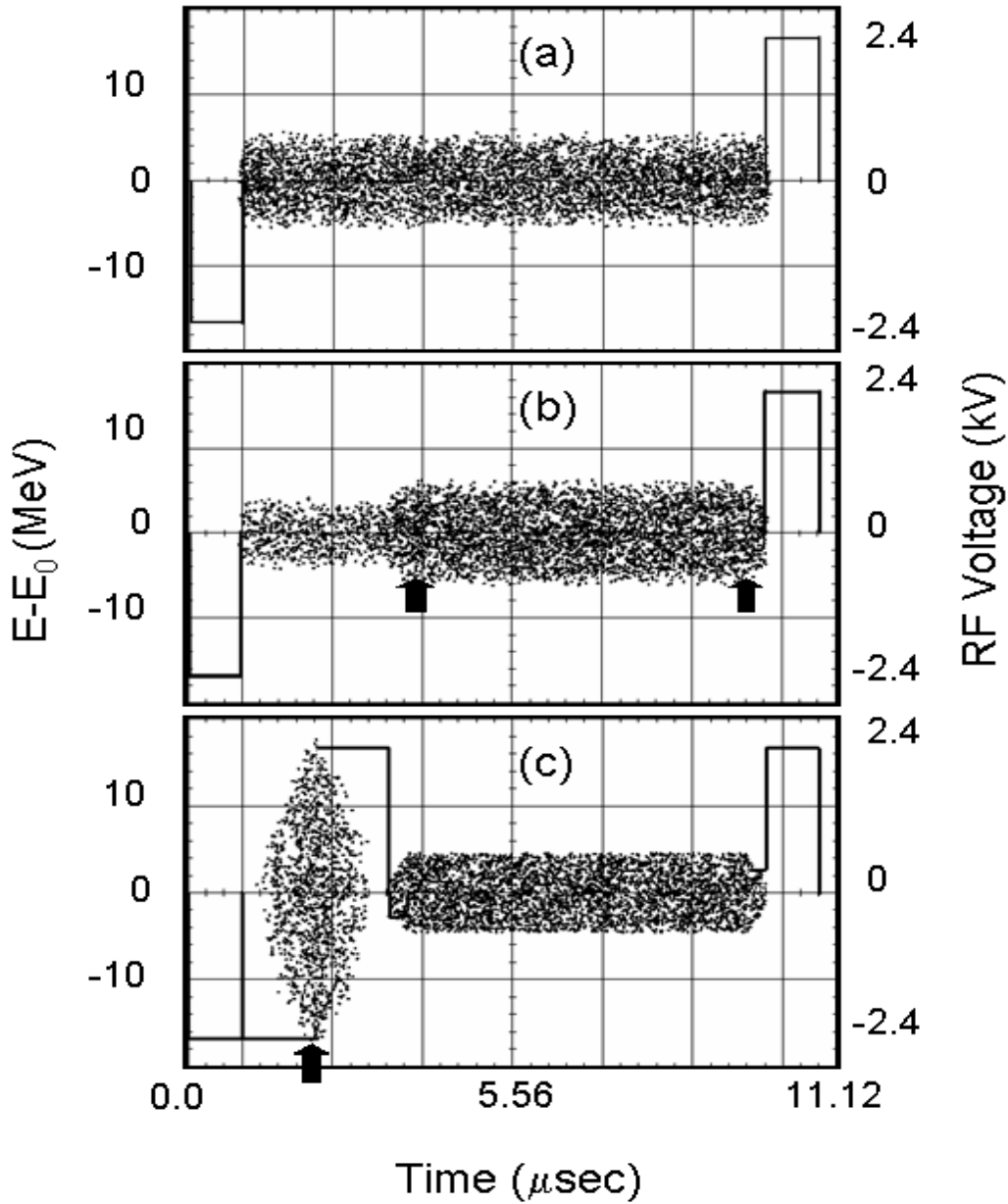

FIG. 2. Simulated ($\Delta E, \tau$) phase-space distribution of 100 eVs anti-protons in the Recycler for (a) initial beam distribution of length 8.7 µsec, (b) after populating the low longitudinal emittance particles to the right hand side using -0.33 kV rf pulse. The arrows indicate the final locations of the rf pulses used. In the simulation, the right side mining rf pulse (+0.33 kV in height) was at a fixed position while the left rf pulse was cogged iso-adiabatically from left to right during shoveling in about 15 sec. (c) After longitudinal momentum mining; the left bunch (indicated by an arrow) comprises of particles with high momentum spread (bunch area ≈51 eVs) and the low longitudinal emittance particles are captured in the right-side bucket (bunch area = 54 eVs).

Fig. 2 shows the beam particle distribution in ($\Delta E, \tau$)-phase space for different stages of mining. The beam distributions before and during mining are shown in Fig. s 2(a) (energy spread of ±5.7 MeV) and 2(b), respectively. The width and amplitude of the barrier pulses used were, respectively, 0.9 μsec and 2 kV for the initial distribution (Fig. 2(a)). Fig. 2(b) shows the distribution after populating all of the low emittance particles to the right hand side in a mining bucket (indicated between two arrows). The size of the mining bucket was chosen to be 54 eVs with rf pulse amplitude of 0.27 kV, of width 0.34 μsec and a pulse gap of 6.13 μsec. All particles with energy spread less than the bucket-height of the mining bucket are confined to the right hand side while the rest move freely throughout the original bucket. (These two cases correspond to the schematic picture shown in Figs. 1(a) and 1(b)).

The isolation of 54 eVs low emittance high density beam was accomplished by opening a bucket (indicated by an arrow in Fig. 2(c)) of an area ≥46 eVs; the area and height of this bucket was chosen to be 74 eVs and ±21.7 MeV, respectively. The Fig. 2(c) shows the final phase space distribution after completion of mining. The total phase-space area was found to be preserved to better than 10% at the end of all rf manipulations. All of the emittance growth was seen in the 74 eVs bucket and occurred during the opening of this bucket.

Finally, the 54 eVs beam was divided into nine bunches of equal intensity and equal longitudinal emittance simply by adiabatic capture (e.g., see Fig. 3(b)) and, each of the nine bunches were further divided into four 2.5 MHz bunches each with 1.5 eVs and were prepared for collider operation.

The simulation clearly showed that the amount of beam mined was a strong function of the energy distribution of particles. For a parabolic distribution, 74% of the beam was mined whereas for Gaussian distribution 64% was mined.

The beam tests were carried out using protons in the Recycler. About $170\times10^{10}$ protons of longitudinal emittance $110\pm15$ eVs (95% emittance) and energy spread of $\pm(14.3\pm0.6)$ MeV were stored in a 3.7 μsec wide rectangular barrier bucket of $\pm 2$ kV pulse height. The beam was stretched slowly to 8.7 μsec and the momentum mining was performed essentially following the sequence studied in the simulation. Experimentally, the entire mining process took about 135 sec while the simulation suggested about 110 sec.

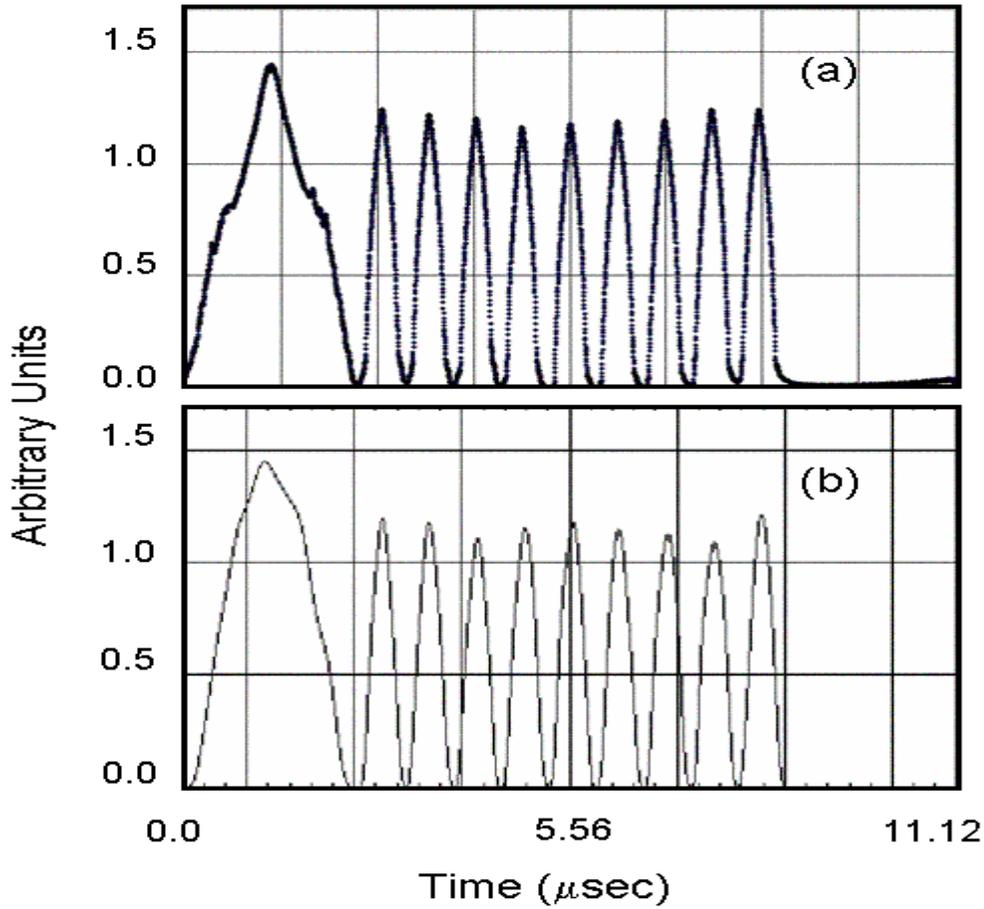

FIG. 3. (a) The measured and (b) predicted line-charge distribution for $170\times10^{10}$ proton after longitudinal momentum mining and capturing high density low longitudinal emittance particles in the nine buckets and the rest in the left most bucket.

The wall current monitor data taken after the formation of nine bunches are shown in Fig. 3(a). The average longitudinal emittance of the beam in these nine buckets is 5.6 eVs ± 0.6 eVs (95% emittance) and the bunch on far left with particles of high momentum spread has longitudinal emittance of 55.5 eVs ± 12.5 eVs (95% emittance). Fig. 3(b) is the corresponding simulated distribution. Experimentally, we find that about 65% of the beam particles are mined in the nine smaller buckets to be compared with 74% predicted by our simulation. The difference is due primarily to the assumed ideal parabolic distribution for the initial energy spectrum, while the un-cooled beam used for the experiments did not resemble a parabolic energy distribution. We also see some qualitative difference between the shapes of the measured wall current monitor data (Fig. 3(a)) and the predictions (Fig. 3(b)). This difference is mainly attributed to (a) the initial energy distribution and (b) the details of rf pulse shapes used in the experiment and the one in the simulation (which was rectangular in shape).

In summary, I have proposed and validated a novel method for selectively isolating low longitudinal emittance particles and extracting of beam bunches from a storage ring using rf barrier buckets. I have studied the scheme using multi-particle beam dynamics simulations and the technique has been demonstrated with beam experiments in the Recycler using protons. This method of momentum mining has been successfully implemented and used routinely for beam extraction from the Fermilab Recycler.

As a final note, I expect that the applications of the technique described here to selectively isolate the high density region of the phase space may not be unique to high-energy storage rings; it should have broad application in other low energy circular storage rings that use barrier rf systems.

I would like to thank John Marriner, Pushpa Bhat, Jim MacLachlan and Harrison Prosper for many useful discussions during the course of this work. Thanks are also due to Brian Chase for his help on issues related to rf controls. My special thanks are due to Shreyas Bhat for a careful editing of the manuscript. I acknowledge the support of U. S.